\documentclass{emulateapj}
\usepackage{natbib}
\usepackage{graphicx}
\newcommand{\be}{\begin{equation}}
\newcommand{\ba}{\begin{eqnarray}}
\newcommand{\ee}{\end{equation}}
\newcommand{\ea}{\end{eqnarray}}

\def\lesssim{\mathrel{\hbox{\rlap{\hbox{\lower4pt\hbox{$\sim$}}}\hbox{$<$}}}}
\def\gtrsim{\mathrel{\hbox{\rlap{\hbox{\lower4pt\hbox{$\sim$}}}\hbox{$>$}}}}

\def\gtsima{$\; \buildrel > \over \sim \;$}
\def\ltsima{$\; \buildrel < \over \sim \;$}
\def\gsim{\lower.5ex\hbox{\gtsima}}
\def\lsim{\lower.5ex\hbox{\ltsima}}
\def\simgt{\lower.5ex\hbox{\gtsima}}
\def\simlt{\lower.5ex\hbox{\ltsima}}
\def\simpr{\lower.5ex\hbox{\prosima}}

\def\simless{\mathbin{\lower 3pt\hbox
   {$\rlap{\raise 5pt\hbox{$\char'074$}}\mathchar''7218$}}}   
\def\simgreat{\mathbin{\lower 3pt\hbox
   {$\rlap{\raise 5pt\hbox{$\char'076$}}\mathchar''7218$}}}   

\begin{document}

\shorttitle{21 cm Background from the Cosmic Dark Ages}
\shortauthors{Shapiro et al.}

\title{The 21 centimeter Background from the Cosmic Dark Ages: Minihalos 
and the Intergalactic Medium before Reionization }

\author{Paul R. Shapiro\altaffilmark{1}, Kyungjin Ahn\altaffilmark{1}, 
Marcelo A. Alvarez\altaffilmark{1}, Ilian T. Iliev\altaffilmark{2},
Hugo Martel\altaffilmark{3}, and Dongsu Ryu\altaffilmark{4} }

\altaffiltext{1}{Department of Astronomy, 1 University Station, C1400,
  Austin, TX 78712, USA}
\altaffiltext{2}{Canadian Institute for Theoretical Astrophysics, University
  of Toronto, 60 St. George Street, Toronto, ON M5S 3H8, Canada}
\altaffiltext{3}{D\'epartement de physique, de g\'enie 
        physique et d'optique,          
        Universit\'e Laval, Qu\'ebec, QC G1K 7P4, Canada
}
\altaffiltext{4}{Department of Astronomy and Space Science,
         Chungnam National University, Daejeon 305--764, Korea
}

\begin{abstract}
The H atoms inside minihalos (i.e. halos with virial temperatures 
$T_{\rm vir} \le 10^{4} {\rm K}$, in the mass range roughly from
$10^{4} M_{\odot}$ to $10^{8} M_{\odot}$) during the cosmic dark ages
in a $\Lambda$CDM universe produce a redshifted background of
collisionally-pumped 21-cm line radiation which can be seen in
emission relative to the cosmic microwave background
(CMB). Previously, we used semi-analytical calculations of the 21-cm
signal from individual halos of different mass and redshift and the
evolving mass function of minihalos to predict the mean brightness
temperature of this 21-cm background and its angular fluctuations.
Here we use high-resolution cosmological N-body and hydrodynamic
simulations of  
structure formation at high redshift ($z\gtrsim 8$) to compute the
mean brightness temperature of this background from
both minihalos and the intergalactic medium (IGM) prior to the onset
of Ly$\alpha$ radiative pumping.
We find that 
the 21-cm signal  from gas in collapsed, virialized minihalos
dominates over that from
the diffuse shocked gas in the IGM.
\end{abstract}


\section{Introduction}
\label{sec:Introduction}

One of the most promising means by which to observe the high redshift
universe in the cosmic ``dark ages'' is through the 21-cm wavelength
hyperfine transition of the neutral hydrogen that is abundant prior
to reionization \citep[e.g.][]{1990MNRAS.247..510S,1993MNRAS.265..101S}.
Motivated by the prospect of new radio telescopes that will be able to 
observe such a signal, several specific observational techniques have been 
proposed. Among these are the angular fluctuations on the
sky (e.g. \citealt*{1997ApJ...475..429M};  \citealt{2000ApJ...528..597T};
\citealt{2002ApJ...572L.123I} -- ISFM hereafter; \citealt{2003ApJ...596....1C,
2003MNRAS.341...81I}; \citealt*{2004ApJ...608..622Z};
\citealt*{2004MNRAS.347..187F}), features in  
the frequency spectrum of the signal averaged over a substantial
patch of the sky \citep{1999A&A...345..380S,2004ApJ...608..611G} and 
studies of absorption features in the spectra of bright, high-redshift
radio sources (\citealt*{2002ApJ...577...22C}; \citealt{2002ApJ...579....1F,2003AIPC..666...85M}).
 
For most of these techniques, with the exception of foreground
absorption against bright radio sources, the 21-cm signal must be
distinguished from the CMB, which is only possible if the 21-cm level
population corresponds to a spin temperature $T_{\rm S}$, which
differs from the temperature, $T_{\rm CMB}$, of the CMB. Since
radiative excitation and stimulated emission of this transition by CMB
photons tends to drive the value of $T_{\rm S}$ toward $T_{\rm CMB}$,
some competing mechanism must exist to decouple $T_{\rm S}$ from
$T_{\rm CMB}$. 
There are two main physical mechanisms 
by which the spin temperature is decoupled from the CMB temperature: 
``Ly$\alpha$ pumping,'' or absorption of radiation with a wavelength
in the Ly$\alpha$ transition, followed by decay into one of the
hyperfine levels of the ground state
(the ``Field-Wouthuysen effect'' -- \citealt{1952AJ.....57R..31W,
1959ApJ...129..536F}), and spin exchange during collisions between neutral 
hydrogen atoms \citep{1956ApJ...124..542P}. 
The efficiency of Ly$\alpha$ pumping depends upon the intensity of the UV 
radiation field at the Ly$\alpha$ transition, whereas the efficiency of 
collisional coupling depends upon the local gas density and temperature. 

For $z\gtrsim 150$, these mechanisms are ineffective at decoupling
$T_{\rm S}$ from $T_{\rm CMB}$, since the kinetic temperature of the
gas, $T_{\rm K}$, is coupled to $T_{\rm CMB}$ by inverse Compton
scattering, and sources of UV radiation have not yet formed to
initiate Ly$\alpha$ pumping. At $z\lesssim 150$, however, $T_{\rm K}$
drops below $T_{\rm CMB}$ and, for $z\gtrsim 20$, gas at the mean
density is sufficiently 
dense for collisions to couple $T_{\rm S}$ to $T_{\rm K}$. During the
``dark ages,'' therefore, when there is no Ly$\alpha$ pumping, the
mean 21-cm signal against the CMB will be zero at $z\gtrsim 150$, then
in absorption at  $20 \lesssim z\lesssim 150$.

At lower redshift, collisions become negligible
for gas at or below the cosmic mean density, and such gas becomes invisible
until its spin temperature is again decoupled from the CMB by Ly$\alpha$
pumping due to an early UV background from the first stars and quasars.
Even though collisional decoupling is ineffective for $z\lesssim 20$
for gas at the {\it mean} density, gas in overdense and/or heated
regions can still 
be collisionally-decoupled. In particular, the gas density within
``minihalos'' -- virialized halos of dark and baryonic matter with 
masses $10^{4} \lesssim M \lesssim 10^{8}M_{\sun}$ and virial
temperatures $T<10^{4}K$ which are too low to collisionally ionize
their H atoms  -- is
sufficiently high so as to decouple its gas spin temperature from the
CMB, with $T_{\rm S} > T_{\rm CMB}$ in general, causing it to
appear in emission (ISFM). ISFM predicted 
the mean and angular fluctuations of the corresponding 21-cm signal by
a semi-analytical calculation which integrated the equation of
transfer through individual minihalos of different mass at different
redshifts ($z > 6$) and summed these individual halo contributions
over the evolving statistical distribution of minihalo masses in the
$\Lambda$CDM universe.
\citet{2003MNRAS.341...81I} extended these
results to include non-linear biasing effects.
These authors concluded that the fluctuations
in intensity across the sky created by minihalos were likely to be
observable by the next generation of low-frequency
radio telescopes. Such observations
could confirm the basic CDM paradigm and constrain the shape and amplitude
of the power spectrum at much smaller scales than previously possible.

Since then, \citet{2004ApJ...611..642F} have suggested that 
shocked, overdense gas in the diffuse IGM (prior to the onset of
Ly$\alpha$ radiative pumping) is also capable of producing a 21-cm
emission signal and that this IGM contribution to the mean signal will
dominate over that from gas inside minihalos.
Their conclusion is based on an extension
of the Press-Schechter approximation (\citealt{1974ApJ...187..425P})
that is used to determine the 
fraction of baryons in the diffuse IGM
that are hot and dense enough
to produce a 21-cm emission signal. We will address this question here.

In order to quantify these effects, we have computed the 21-cm signal
both from minihalos and the IGM at $z\gtrsim 8$ for the first time
using high-resolution cosmological N-body and hydrodynamic simulations
of structure formation. We have assumed a flat, $\Lambda$CDM cosmology
with matter density parameter  $\Omega_{m}=0.27$, cosmological constant 
$\Omega_{\Lambda}=0.73$, baryon density $\Omega_b=0.043$, Hubble constant 
$H={\rm 70\, km\,s^{-1}Mpc^{-1}}$, linearly-extrapolated
$\sigma_{8h^{-1}}=0.9$ and the ``untilted''
Harrison-Zel'dovich primordial power spectrum.

In this paper, we present detailed, high-resolution gas and N-body
simulations which
predict the 21-cm signal at $z>6$ due to collisional
decoupling from the CMB before the UV background is strong enough to make 
decoupling due to Ly$\alpha$ pumping important. Because the Ly$\alpha$ 
pumping efficiency is expected to fluctuate strongly until enough sources 
form to make the efficiency uniform \citep[e.g.][]{2004ApJ...609..474B}, 
the results presented here will also be relevant for isolated patches of the
universe during reionization itself, which would depend upon the location and
abundance of the first sources of UV radiation. Within such regions,
we focus on properly resolving the gasdynamics of structure formation
at small scales through the use of high resolution gasdynamic and
N-body simulations. We test the semi-analytical prediction of the halo 
model of ISFM for the contribution to the mean signal from gas in minihalos, 
and investigate the extent to which IGM gas may provide a non-negligible 
contribution to the total fluctuating signal, as suggested by 
\citet{2004ApJ...611..642F}. 

These results were first summarized in
\citet{2006NewAR..50..179A}. Here we shall describe our calculations
in full and present our results in more detail.

The outline of this paper is as follows. 
In \S~\ref{sec:Calculation} we summarize the basic physics of the 21-cm
emission and absorption and the analytical model of ISFM. We also describe 
our cosmological simulations and their initial conditions, and our method 
for obtaining the 21-cm signal from our simulations. In \S~\ref{sec:Result} 
we present our results.
Our conclusions are summarized in \S~\ref{sec:Conclusion}.

\section{The Calculation}
\label{sec:Calculation}

\subsection{Physics of 21-cm signal from neutral hydrogen}
\label{sub:basic}

The hyperfine splitting of the ground state of hydrogen leads to a
transition with excitation temperature $T_{*}=0.068 \,{\rm K}$,
wavelength $\lambda_{0}=21.16 \,{\rm cm}$, and frequency $\nu_{0}=1.417
\,{\rm GHz}$. The ratio of populations of the upper ($n_1$) and lower
($n_0$) states is characterized by the spin temperature $T_{\rm S}$
according to 
\begin{equation}
\label{eq-ts-def}
\frac{n_1}{n_0}=3 \exp\left( -T_*/T_{\rm S}\right).
\end{equation}
Neutral hydrogen at redshift $z$ produces a differential signal
relative to the CMB at redshifted wavelength $21(1+z)\,
{\rm cm}$ only if $T_{\rm S}$ differs from $T_{\rm CMB}$. The $21
\,{\rm cm}$ transition is seen in emission against the CMB if $T_{\rm
  S} > T_{\rm CMB}$ or in absorption if $T_{\rm
  S} < T_{\rm CMB}$. The value of $T_{\rm S}$ is determined by the
relative importance of collisional and radiative excitations.
A hydrogen atom can 1) absorb a 21-cm photon from the CMB (CMB pumping), 2)
collide with another atom (collisional pumping) and 3) absorb a
Ly$\alpha$ photon to make a Ly$\alpha$ transition, then decay to one of
the hyperfine 21-cm levels (Ly$\alpha$ pumping).  
These pumping mechanisms
jointly determine the spin temperature,
\begin{equation}
\label{eq-ts}
T_{S}=\frac{T_{{\rm
      CMB}}+y_{\alpha}T_{\alpha}+y_{\rm c}T_{\rm K}}{1+y_{\alpha}+y_{\rm c}},
\end{equation}
 where $T_{\alpha}$ is the
color temperature of the Lyman-$\alpha$ photons, $T_{\rm K}$ is the
kinetic temperature, $y_{\alpha}$ is the Lyman-$\alpha$ coupling
constant, and $y_{\rm c}$ is the collisional coupling constant 
\citep{1956ApJ...124..542P,1959ApJ...129..536F}.
As seen in equation~(\ref{eq-ts}), the spin temperature
deviates from $T_{\rm CMB}$ only when these couplings exist. Throughout
this paper, we will consider only the collisionally coupled gas, or
the case where $y_{\alpha}=0$. This is valid when 1) 
light sources were not yet abundant enough to build up substantial
Lyman-$\alpha$ background or 2) the region
of interest is far enough away from light sources. 
The collisional coupling constant is given by
\begin{equation}
\label{eq-coll-coup}
y_{\rm c} = \frac{T_{*}C_{10}}{T_{\rm K}A_{10}},
\end{equation}
where $A_{10}=2.85\times 10^{-15} {\rm s}^{-1}$ is
the Einstein 
spontaneous emission coefficient, and $C_{10}=\kappa(1-0)n_{\rm H}$ is the
atom-atom collisional de-excitation rate \citep{1956ApJ...124..542P}. We
use $\kappa(1-0)$ tabulated in 
\citet{2005ApJ...622.1356Z} which is valid for $1 {\rm K} < T_{\rm K}
< 300 {\rm K}$, and for higher $T_{\rm K}$ we use $\kappa(1-0)$
tabulated in \citet{1969ApJ...158..423A}\footnote{Definition of
  $\kappa(1-0)$ in 
  \citet{1969ApJ...158..423A} is not consistent with that in
  \citet{2005ApJ...622.1356Z}. One should multiply 
  their $\kappa(1-0)$ by $(4/3)$ in order to calculate $C_{10}$ when using
  tabulated $\kappa(1-0)$ of \citet{1969ApJ...158..423A}.
}.

The 21-cm line can be observed in either absorption or emission
against the CMB, 
with a differential brightness temperature defined by 
\begin{equation}
\label{eq-dT-def}
\delta T_{b} (\nu)\equiv T_{b} (\nu) - T_{\rm CMB, 0} ,
\end{equation}
where $T_{b} (\nu)$ is the brightness temperature at an observed
frequency $\nu$ and $T_{\rm CMB, 0}$ is the present-day CMB
temperature. $T_{b} (\nu)$ satisfies the radiative transfer equation
in the Rayleigh-Jeans limit,
\begin{equation}
\label{eq-rad-trans}
T_{b} (\nu) = T_{\rm CMB, 0} e^{-\tau_{\nu}}
+ \int_{0}^{\tau_{\nu}} d\tau'_{\nu'} \frac{T_{\rm
S}(z')}{1+z'} e^{-(\tau_{\nu}-\tau'_{\nu'})} ,
\end{equation}
where $\tau_{\nu}=\int_{0}^{\tau_{\nu}} d\tau'_{\nu'}$ is the 21-cm
optical depth of the neutral hydrogen atoms to photons in the CMB 
observed today at frequency $\nu$, 
$\tau'_{\nu'}=\int_{0}^{\tau'_{\nu'}} d\tau''_{\nu''}$ is
the optical depth of the neutral hydrogen atoms 
at redshift $z'$ to photons at
a frequency $\nu'=\nu (1+z') $ (the frequency which a comoving
observer sees at redshift $z'$),
$T_{\rm S}(z')$ is the spin temperature of intervening gas located at
redshift $z'$, 
and the infinitesimal optical depth along the path of the photon as it
travels for cosmic time interval $dt'$ is given by
\begin{eqnarray}
d \tau'_{{\nu'}} &=& c \, dt' \,\kappa_{\nu'}(\nu',z') \nonumber \\
&=&\left[ \frac{c\,dz'}{H(z')(1+z')} \right]
\left[ \frac{3c^{2}A_{10}n_{\rm HI}(z')}{32\pi {\nu'}^{2}}
\phi(\nu') \frac{T_{*}}{T_{\rm S}(z')} \right],
\end{eqnarray}
where $H(z)$ is the Hubble constant at redshift $z$,
and $n_{\rm HI}$ is the local density of neutral hydrogen.
The line profile $\phi(\nu')$ satisfies
\begin{equation}
\label{eq-phi}
\int_{-\infty}^{+\infty}d\nu' \phi(\nu') =1,
\end{equation}
and is in a general form. For instance, in the presence of both thermal
Doppler broadening and the Doppler shift
due to peculiar motion, the line profile is given by
\begin{equation}
\label{eq-general-phi}
\phi(\nu')=\frac{1}{\Delta \nu' \sqrt{\pi}}
\exp\left[
-\frac{(\nu'-\nu'_{0})^2}{{\Delta \nu'}^2}
\right],
\end{equation}
where $\Delta \nu' = (\nu'_{0}/c) \sqrt{2kT/m}$, and the Doppler-shifted
line center is given by
\begin{equation}
\nu'_{0} = \nu_{0} \frac{1+\beta}{\sqrt{1-\beta^2}},
\end{equation}
where $\beta={\rm v}/c$ is the line-of-sight peculiar velocity of gas (in
units of the speed of light) {\it toward} the observer.

\subsubsection{The unperturbed universe}
\label{sub:unperturbed}

\begin{figure}
\plotone{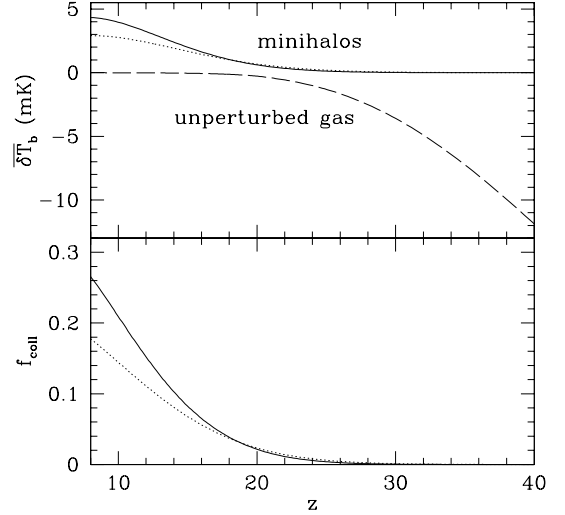}
\caption{Analytical prediction for the mean 21-cm 
differential brightness temperature due to collisionally-decoupled 
minihalos and an unperturbed IGM. Shown are the results based on the 
Press-Schechter (solid) and the Sheth-Tormen (dotted) mass functions 
for halos and the contribution from IGM gas with cosmic mean 
density and temperature (dashed). In the bottom panel, we show the 
minihalo collapsed fraction in the $\Lambda$CDM universe, again based
on the Press-Schechter (solid) and  
the Sheth-Tormen (dotted) mass functions.}
\label{fig-anal}
\end{figure}

Solutions to the general radiative transfer equation
(eq. [\ref{eq-rad-trans}]) exist in simplified forms in limiting
cases. If the line is un-broadened and
un-shifted, i.e. $\phi(\nu') = \delta(\nu'-\nu_{0})$, the solution
to equation (\ref{eq-rad-trans}) becomes
\begin{equation}
\label{eq-sol}
T_b (\nu) = T_{\rm CMB, 0} e^{-\tau(z)}
+ \frac{T_{\rm S}(z)}{1+z} \left[1-e^{-\tau(z)}\right],
\end{equation}
where $\nu$ and $z$ satisfy $\nu_{0}=\nu (1+z)$, and
\begin{eqnarray}
\label{eq-tau}
&&\tau(z) \equiv 
\frac{3\lambda_{0}^{3}A_{10}T_{*}n_{\rm HI}(z)}{32\pi
  T_{S}H(z)} = 3.22\times 10^{-3}  \nonumber \\
&& \times \left( \frac{\Omega_{b}h^2}{0.0224} \right)
\left( \frac{\Omega_{0}h^2}{0.135} \right)^{-0.5}
\left( \frac{T_{\rm CMB}}{T_{\rm S}}\right)
(1+z)^{0.5},
\end{eqnarray}
using $n_{\rm HI}=x_{\rm HI}n_{\rm H}$ , where
$n_{\rm H}= 1.9\times 10^{-7} {\rm cm}^{-3} (1+z)^{3}$
and $x_{\rm HI}$ is the neutral fraction of hydrogen.
The IGM kinetic temperature, $T_{\rm K}$, is coupled to
$T_{\rm CMB}$ by 
Compton scattering at $z \gtrsim 134$. 
For $z \lesssim 100$,
the kinetic temperature of the unperturbed gas in the $\Lambda$CDM
universe is well approximated by
\begin{eqnarray}
\label{eq-Tk-IGM}
T_{\rm K} & \approx & T_{\rm CMB} (z=134)
{(1+z)^2}/{(1+134)^2} \nonumber \\
&=& 368.55 {\rm K} \times {(1+z)^2}/{(1+134)^2}.
\end{eqnarray}
Since $T_{\rm CMB}/T_{\rm S} < {\rm max}\{1,\,T_{\rm CMB}/T_{\rm K}\}$, in
general, $\tau_{\nu} \ll 1$ for the unperturbed IGM 
at all redshifts of interest ($ 8 \lesssim z \lesssim 50 $), as seen
in equation (\ref{eq-tau}). In that case, equation 
(\ref{eq-sol}) can be approximated as 
\begin{equation}
\label{eq-sol-thin}
\delta T_b(\nu) = \left[ T_{\rm S}(z)-T_{\rm CMB}(z) \right]
\tau(z) / (1+z)
\end{equation}
using equation (\ref{eq-dT-def}).
We can use equations (\ref{eq-tau}) and (\ref{eq-sol-thin}) to describe
the signal from the unperturbed gas in the Universe, because the line
profile in this case is narrow enough to be approximated by a
Dirac-$\delta$ function. With equations (\ref{eq-ts}),
(\ref{eq-coll-coup}), (\ref{eq-tau}) and (\ref{eq-sol-thin}), one can
calculate $\delta T_b$ of the unperturbed gas of the Universe,
well approximated by
\begin{equation}
\label{eq-dT-fit}
\delta T_b(z) \approx -10\,{\rm mK}
\frac{F[\log_{10} (1+z)]}{F[\log_{10}(1+36.8)]} \, ,
\end{equation}
where 
\begin{eqnarray}
\label{eq-Fx}
&& F(x) = {\rm dex} (3172.36 - 18037.3 \, x + 43430.3 \, x^2 - 57481.7 \, x^3
\nonumber \\ 
&& + 45150.0 \, x^4 -
21042.1 \, x^5 + 5388.50 \, x^6 - 585.165 \, x^7 ) ,
\end{eqnarray}
which is in absorption until collisional pumping becomes negligible at
$z \simeq 20$ (Fig.~\ref{fig-anal}; see also 
\citealt{2004MNRAS.352..142B} and \citealt{2005ApJ...622.1356Z}). 

\subsubsection{Perturbed universe: optically thin case}
\label{sub:optthin}

\begin{figure*}
\plotone{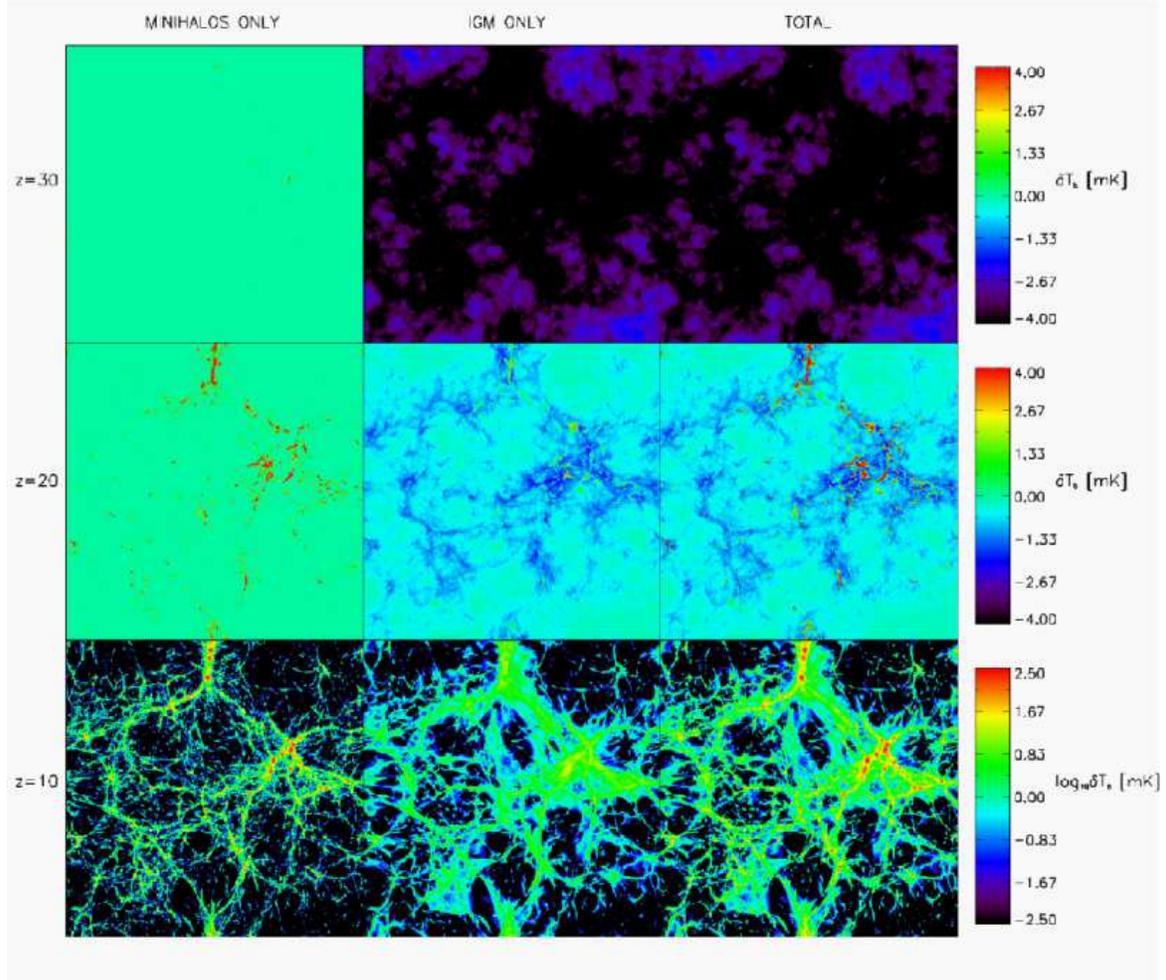}
\caption{Map of the differential brightness temperature,
  $\delta T_{b}$, (projected onto one surface of the box) for the redshifted
21-cm signal obtained from our highest
  resolution simulation, C4. Rows, top to bottom, show redshifts $z$=30, 20,
  and 10. Columns, left to right, represent contributions from minihalos, the
  IGM and the total signal. Note that the scale is linear in
  $\delta T_{b}$ for the upper two 
  rows of images, but logarithmic for the bottom row. 
}
\label{maps}
\end{figure*}

Thermal Doppler
broadening and Doppler shift by peculiar motions would drive $\phi$
to be broadened and shifted, causing overlap of line profiles. In
such cases, the solution to equation (\ref{eq-rad-trans}), in
general, is not given in a simple form as in equation (\ref{eq-sol}).
We show here, however, that the simple solution given by equation
(\ref{eq-sol-thin}) also applies to the overlapped line profile case, as long
as the optical depths, both infinitesimal and integrated, of gas in
the simulation box are small. In such optically-thin limit, equation
(\ref{eq-rad-trans}) can be approximated as
\begin{equation}
\label{eq-rad-trans-thin}
T_{b} (\nu) = T_{\rm CMB, 0} (1-\tau_\nu)
+ \int_{0}^{\tau_{\nu}} d\tau'_{\nu'} \frac{T_{\rm
S}(z')}{1+z'} .
\end{equation}
The differential brightness temperature
from a simulation box 
at $z$ with a redshift-spread $\Delta z$ ($\ll z$) and
angle-spread $\Delta \Omega$ is 
\begin{equation}
\label{eq-freq-angle-int}
\overline{\delta T}_b(\nu) \equiv \frac{\int d\nu d\Omega \, \delta
T_b(\nu)}{\int d\nu d\Omega },
\end{equation}
where the frequency and angle intervals of integration are set by
the size of the box. As the angle integration is a simple sum of different
line-of-sight contributions which do not interfere with each other, we
can first perform the line-of-sight average,
\begin{equation}
\label{eq-freq-int}
\left. \overline{\delta T}_b(\nu) \right|_{\rm l.o.s.} =
\frac{\int^{\nu+\delta \nu/2}_{\nu-\delta \nu/2} d\nu  \, \delta
  T_b(\nu)}{\int^{\nu+\delta \nu/2}_{\nu-\delta \nu/2} d\nu}, 
\end{equation}
and then integrate over angles. In equation (\ref{eq-freq-int}), one
can show that
\begin{equation}
\label{eq-intdnu}
\int^{\nu+\delta \nu/2}_{\nu-\delta \nu/2} d\nu =
\frac{\nu_{0}}{(1+z)^2}\int^{z+\Delta z/2}_{z-\Delta z/2} dz', 
\end{equation}
and
\begin{eqnarray}
\label{eq-intdnudtau}
&&\int^{\nu+\delta \nu/2}_{\nu-\delta \nu/2} d\nu 
\int^{\tau_\nu}_{0} d\tau'_{\nu'} \nonumber \\
&&= \frac{c \nu_{0}^{-2}}{H(z)(1+z)^2} 
\int^{z+\Delta z/2}_{z-\Delta z/2} dz'
\frac{3c^2 A_{10} n_{\rm HI}(z') T_{*}}{32\pi T_{\rm S}(z')},
\end{eqnarray}
where we have used the fact that $\int_{-\infty}^{\infty} d\nu'
\phi(\nu') = 1$, and have also assumed that the thermal broadening and the
Doppler shift by peculiar motions are negligible compared to the width
of the box: $(\Delta \nu')_{\rm thermal} \ll \Delta \nu'_{\rm box}$, 
$(\Delta \nu')_{\rm peculiar} \ll \Delta \nu'_{\rm box}$. Using
equations (\ref{eq-rad-trans-thin}), (\ref{eq-intdnu}), and
(\ref{eq-intdnudtau}), we then obtain 
\begin{equation}
\label{eq-proof}
\left. \overline{\delta T}_b(\nu) \right|_{\rm l.o.s.} =
\int^{\Delta z} dz' \left(
\frac{T_{\rm S(z')}}{1+z'}-T_{\rm CMB,0}  \right) \tau(z')
\bigg/\int^{\Delta z} dz',
\end{equation}
which is simply an averaged superposition of contributions of gas given by
equation (\ref{eq-sol-thin}), along the line-of-sight.

We have shown, in this section, that the 21-cm differential brightness
temperature $\overline{\delta T}_{b}(\nu)$ can be calculated by a
simple average of individual contributions from gas at different
locations in a simulation box, as long as optical depths are small
(eq. [\ref{eq-proof}]). Care needs to be taken, however, when gas
achieves considerable optical depth. In
\S~\ref{sub:minihalos}, we describe how one can calculate
the signal from optically-thick media, which are mostly located inside
minihalos.

\begin{figure*}[t]
\plotone{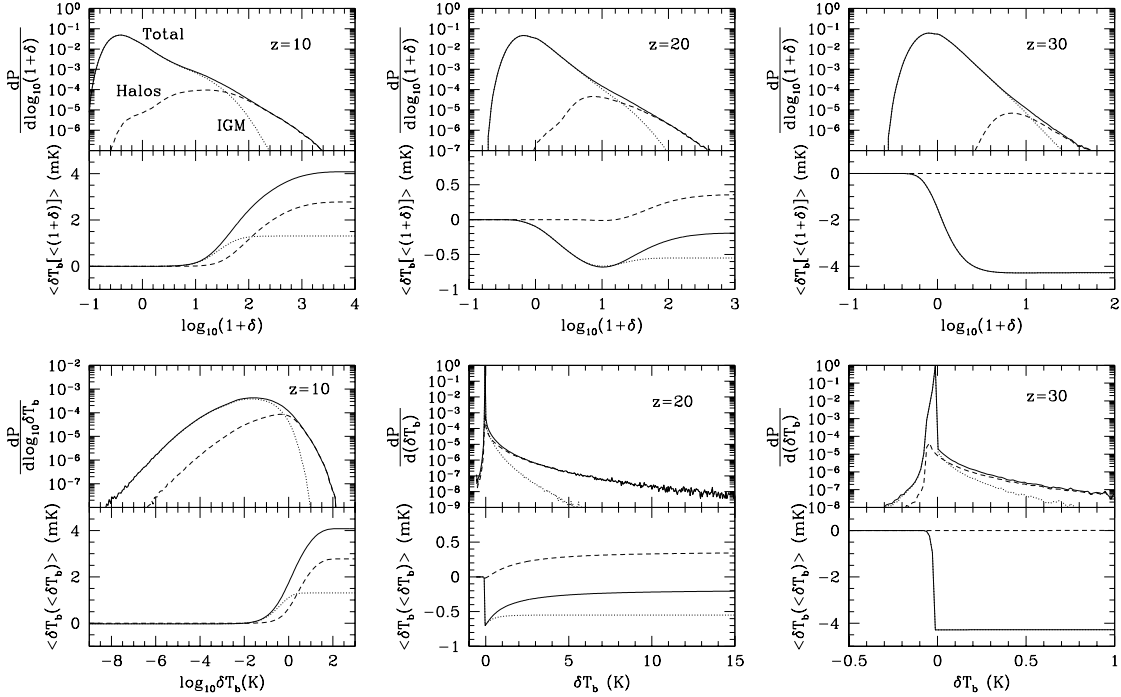}
\caption{
Volume-weighted
probability distribution functions (PDFs)
  of gas  density
  ($1+\delta$) and differential brightness temperature ($\delta
  T_b$) for C4 versus $(1+\delta)$ and $\delta T_b$, respectively, for
  the total gas (solid), MH 
  gas (dashed), and IGM gas (dotted).  Also shown are the
  cumulative  
  differential brightness temperatures, i.e. $\left<\delta T_b
    (<[1+\delta])\right>$ and 
  $\left<\delta T_b (< \delta T_b)\right>$. The top and bottom panels
  from left to right 
  correspond to $z$=10, 20, and 30, respectively.
}
\label{fig-accudelT}
\end{figure*}

\subsubsection{Perturbed universe: minihalos}
\label{sub:minihalos}
Minihalos which start to emerge at $z \simeq 20$
have temperature and density high enough to produce 
a significant emission
signal (ISFM). As the optical depth through each minihalo is not
negligible, the full radiative transfer equation
(eq. [\ref{eq-rad-trans}]) should be solved through 
individual minihalos. Once 
individual halo contribution
$\Delta \nu_{\rm eff}\delta T_{b, \nu_{0}}$ is obtained for each given
halo mass $M$, 
one can calculate  
$\overline{\delta T}_{b}$ from all the minihalos at redshift $z$ by
integrating over the halo mass function $dn/dM$:
\begin{equation}
\label{eq-dTb}
\overline{\delta T}_{b}=\frac{c(1+z)^{4}}{\nu_{0}H(z)}\int_{M_{{\rm
      min}}}^{M_{{\rm max}}}\Delta\nu_{{\rm eff}}\delta
T_{b,\,\nu_{0}}A\frac{dn}{dM}dM,
\end{equation}
where $\Delta\nu_{{\rm eff}}$, $\delta T_{b,\,\nu_{0}}$, and $A$ 
refer to the parameters of the individual virialized halo, 
$\Delta\nu_{{\rm eff}}=\Delta\nu'(1+z)^{-1}=(\nu'_{0}/c)
\sqrt{2kT/m}(1+z)^{-1}$, $\delta T_{b,\,\nu_{0}}$ is the face-averaged
differential brightness temperature at line center, and $A$ is the
projected surface 
area of the halo. ISFM based their calculation on the nonsingular,
truncated isothermal sphere (TIS) model for CDM halos by 
 \citet*{1999MNRAS.307..203S} and \citet{2001MNRAS.325..468I}, in which
 the halo density profile and virial temperature are fully specified
 by the halo mass and redshift, and the Press-Schechter (1974) mass
   function, which determines the number density of halos at a given redshift.
The minimum minihalo mass  
$M_{{\rm min}}$ is set by the Jeans mass
\begin{equation}
\label{eq-jeans}
M_{{\rm J}}=
5.7\times10^{3}\left(\frac{\Omega_{m}h^{2}}{0.15}\right)^{-1/2}
\left(\frac{\Omega_{b}h^{2}}{0.02}\right)^{-3/5}
\left(\frac{1+z}{10}\right)^{3/2}M_{\sun}.
\end{equation}
For $M_{{\rm max}}$, ISFM used the mass for which the virial temperature
is $10^{4}{\rm K}$: 
\begin{equation}
\label{eq-mmax}
M_{{\rm max}}=
3.95\times10^{7}\left(\frac{\Omega_{m}h^{2}}{0.15}\right)^{-1/2}
\left(\frac{1+z}{10}\right)^{-3/2}M_{\sun}.
\end{equation}
The neutral baryonic fraction of halos with mass above $M_{{\rm max}}$ is
uncertain, because hydrogen will be partially ionized due to collisions and 
photoionization by internal sources. Thus, the mass range from $M_{{\rm min}}$ 
to $M_{{\rm max}}$ naturally defines the mass range of minihalos which
are fully neutral.
Figure~\ref{fig-anal} depicts the predicted signals from unperturbed
gas as well as from minihalos. We show results for both the
Press-Schechter and the Sheth-Tormen mass functions
(\citealt{2002MNRAS.329...61S}).

\subsection{Numerical Simulations}
\label{sec:Numerical-Simulations}

We have run series of cosmological N-body and gasdynamic simulations to derive
the effect of gravitational collapse and the hydrodynamics on the predicted 21
cm signal from high redshift. Our computational box has a comoving size of 
$0.5\, h^{-1}\,{\rm Mpc}$, which is optimal for adequately resolving both the
minihalos and the small-scale structure-formation shocks. We used the code
described in \cite{1993ApJ...414....1R}, which uses the particle-mesh (PM) scheme
for calculating the gravity evolution and an Eulerian total variation diminishing
(TVD) scheme for hydrodynamics. We generated our initial conditions for the
gas and dark matter distributions using the publicly available  software COSMICS
\citep{1995ApJ...455....7M}. The N-body/hydro code uses an $N^{3}$ grid and 
$(N/2)^{3}$ dark matter particles. In order to check the convergence of our
results we ran simulations at different spatial
resolutions, with grid sizes $128^{3}$, $256^{3}$, $512^{3}$, and $1024^{3}$,
which we denote by C1, C2, C3, and C4, respectively. We report our results in
\S~\ref{sec:Result} based on our highest-resolution simulation C4 and discuss
the convergence of the results in \S~\ref{conv_sect}.

\begin{figure*}[t]

\plottwo{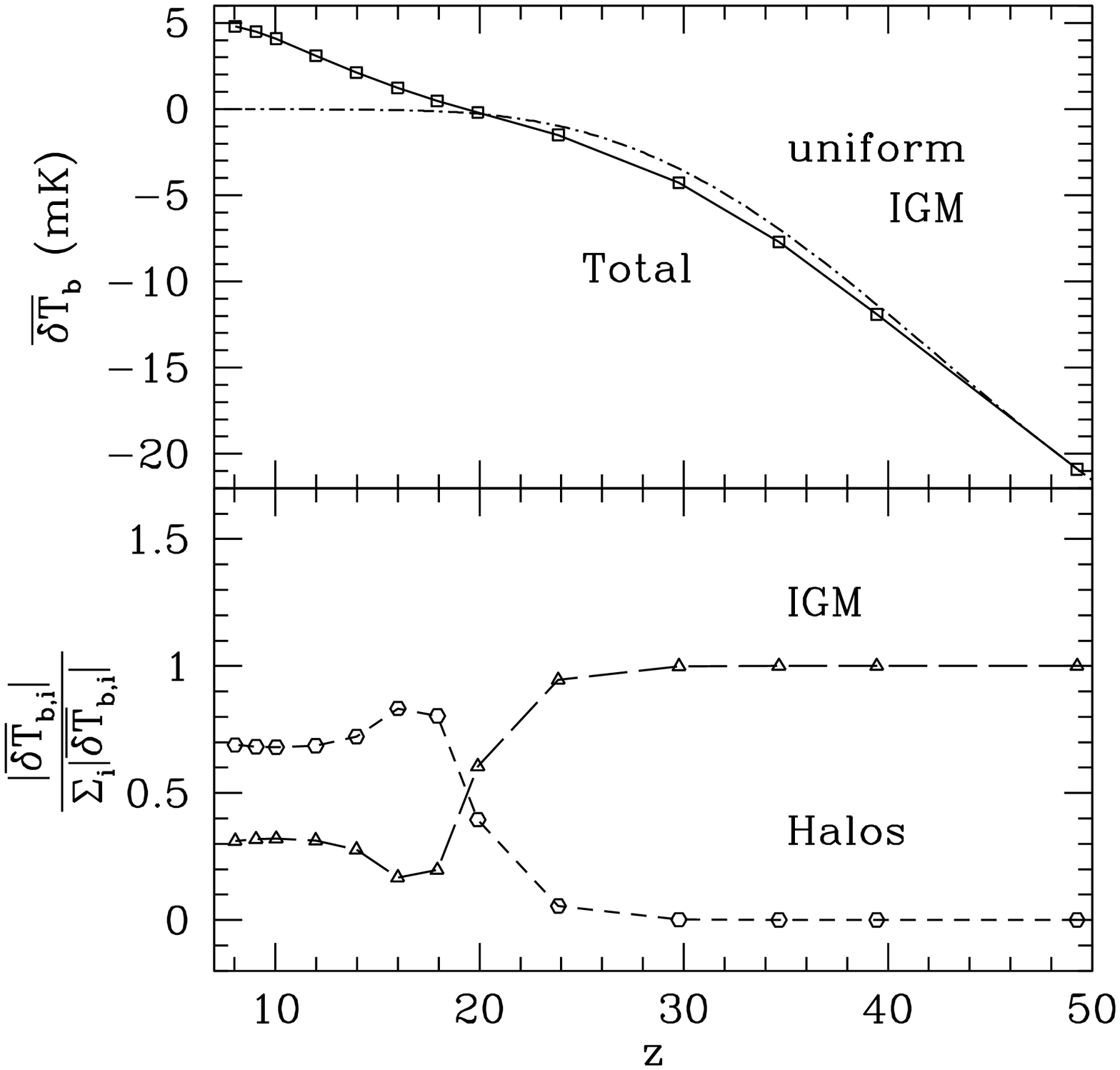}{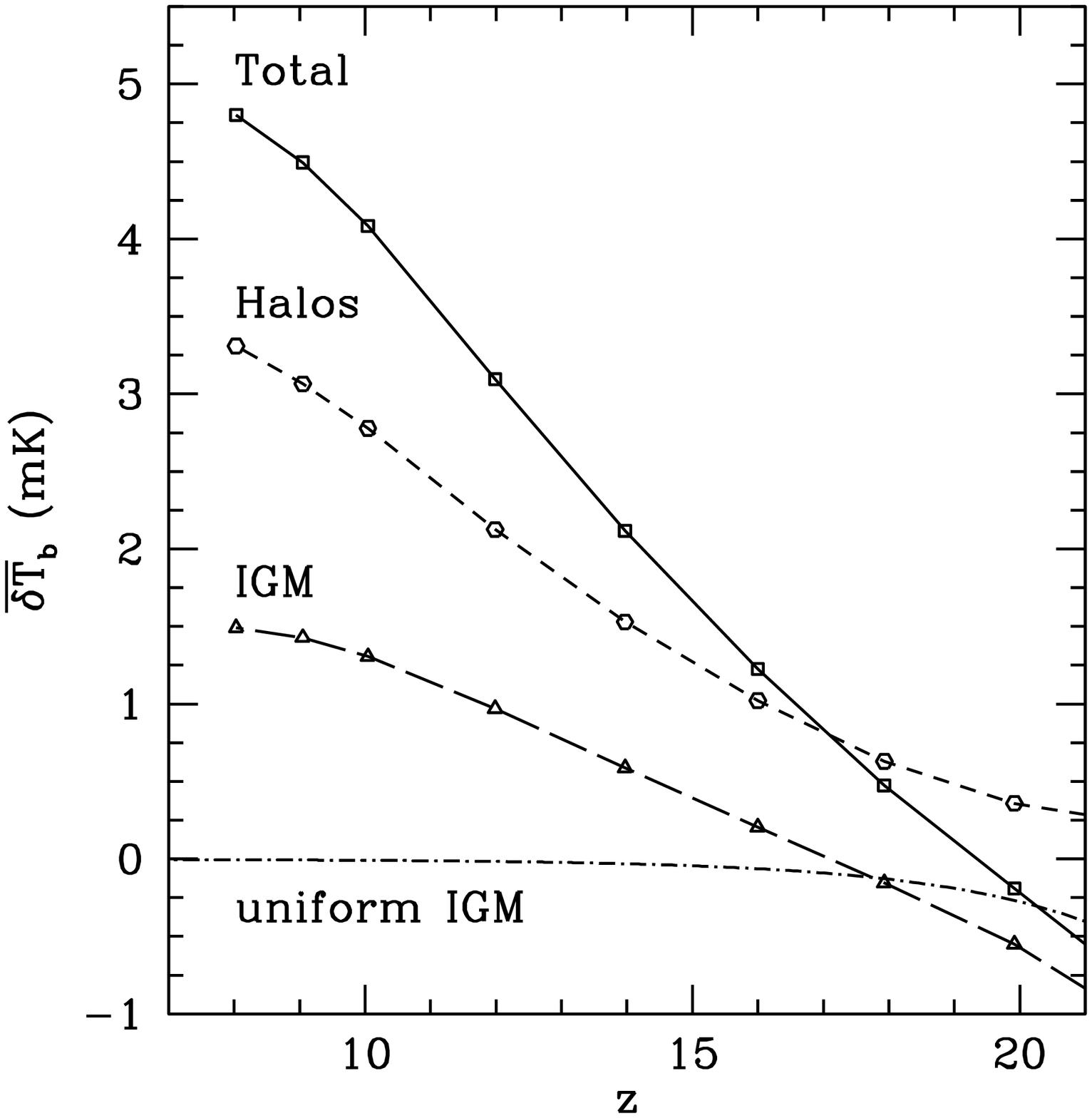}

\caption{
Evolution of
mean differential brightness temperature, $\overline{\delta T}_b$, of
21-cm background. 
(a)(left) Evolution of the total 21-cm signal vs. redshift. 
All data points are directly calculated from our 
highest resolution (C4) simulation box,
with the assumption that optical depth is negligible throughout the box. 
(b)(right) $\overline{\delta T}_b$ vs. redshift below $z=20$. 
The
contributions from minihalos (circles), the IGM (triangles),
and the total (squares) are plotted, as labelled. For comparison, the
result for the unperturbed IGM is also plotted (dashed-dot curves).}
\label{fig-0.25-numerical}
\end{figure*}

After the decoupling of CMB photons from the baryonic gas, the IGM gas cools
adiabatically due to cosmic expansion (eq. [\ref{eq-Tk-IGM}]). 
Equation (\ref{eq-Tk-IGM}) agrees, for instance, with
the solution to 
the equation (1) in \citet{2004MNRAS.352..142B} which describes how
$T_{\rm gas}$ evolves exactly.
This temperature, $T_{{\rm gas}}(z)$, was used in the simulation to set
the minimum temperature of baryonic gas, to avoid negative
temperatures\footnote{One should, in principle, use the locally varying 
minimum temperature. However, usage of a global minimum temperature is well 
justified as described in the text, and it is computationally cheaper than 
implementing a locally varying minimum temperature.}. If a gas cell is cooled 
below $T_{{\rm gas}}(z)$, its temperature is set back to $T_{{\rm gas}}(z)$.
Such a temperature ``floor'' may overestimate the gas temperature
of underdense regions, but because of their low density and temperature,
$y_{c}$ is small in these regions. This implies that
the spin temperature $T_{\rm S}$ would be very close
to $T_{\rm CMB}$, and their contribution to $\delta T_{b}$ would also
be negligible, whether the kinetic temperature $T_{\rm K}$ is calculated
accurately or not.

In addition to the total 21-cm signal from our simulations, 
$\overline{\delta T}_{b,\,{\rm IGM}}$, we are also interested in the relative
contribution of the virialized minihalos and  the IGM to
the total signal, the sum of which gives the total 21-cm signal,  
$\overline{\delta T}_{b,\,{\rm tot}}=
\overline{\delta T}_{b,\,{\rm halo}}+\overline{\delta T}_{b,\,{\rm IGM}}$. 
First, we calculate the total mean signal as a simple average over the simulation
cells, $\overline{\delta T}_{b,\,{\rm tot}}
            \equiv{ \sum_{i}\delta T_{b,\, i}}/N^3$.
The minihalo contribution is given by 
$\overline{\delta T}_{b,\,{\rm halo}}
\equiv{ \sum_{i}f_{i}\delta T_{b,\, i}}/N^3$,
where $f_{i}$ is the fraction of the DM mass in a cell $i$
which is part of a halo.  
The IGM contribution can then be obtained as 
\begin{equation}
\overline{\delta T}_{b,\,{\rm IGM}}
=\overline{\delta T}_{b,\,{\rm tot}}-\overline{\delta T}_{b,\,{\rm halo}}
={\displaystyle \sum_{i}(1-f_{i})\delta T_{b,\, i}}/N^3.
\end{equation} 

In order to calculate the minihalo contribution to the total differential
brightness temperature, $\overline{\delta T}_{b,\,{\rm halo}}$, one needs
to first identify the halos in the simulation volume. We identified the halos 
using a friends-of-friends (FOF) algorithm \citep{1985ApJ...292..371D} with a
linking length parameter of $b=0.25$. 
The FOF algorithm applies to the dark matter N-body particles, rather
than the gas in grid cells. Once this halo catalogue is processed for
each time-slice of our N-body results, the baryonic component of each
halo is identified for the grid cells of the hydrodynamics simulation
which are contained within the volume of the halos in our FOF
catalogue. We do this as follows. First, the density in each cell
contributed by each DM particle is determined by the
triangular-shaped cloud assignment scheme.
For each cell in which mass is contributed by the DM particles
of a given halo, the gaseous baryonic component in that cell is
assumed to contribute a fraction $f_{i}$ of its mass given by the
fraction of the total DM mass in that cell which is attributed to the
halo DM particles. Accordingly, each cell $i$ contributes an amount
$f_{i} \delta T_{{b},i}$ to the signal attributed to halo gas,
while $(1-f_{i})\delta T_{{b},i}$ is assumed to be the signal from
the IGM outside of the halo, where $\delta T_{{b},i}$ is
calculated from the cell as a whole.

\begin{figure*}
\plottwo{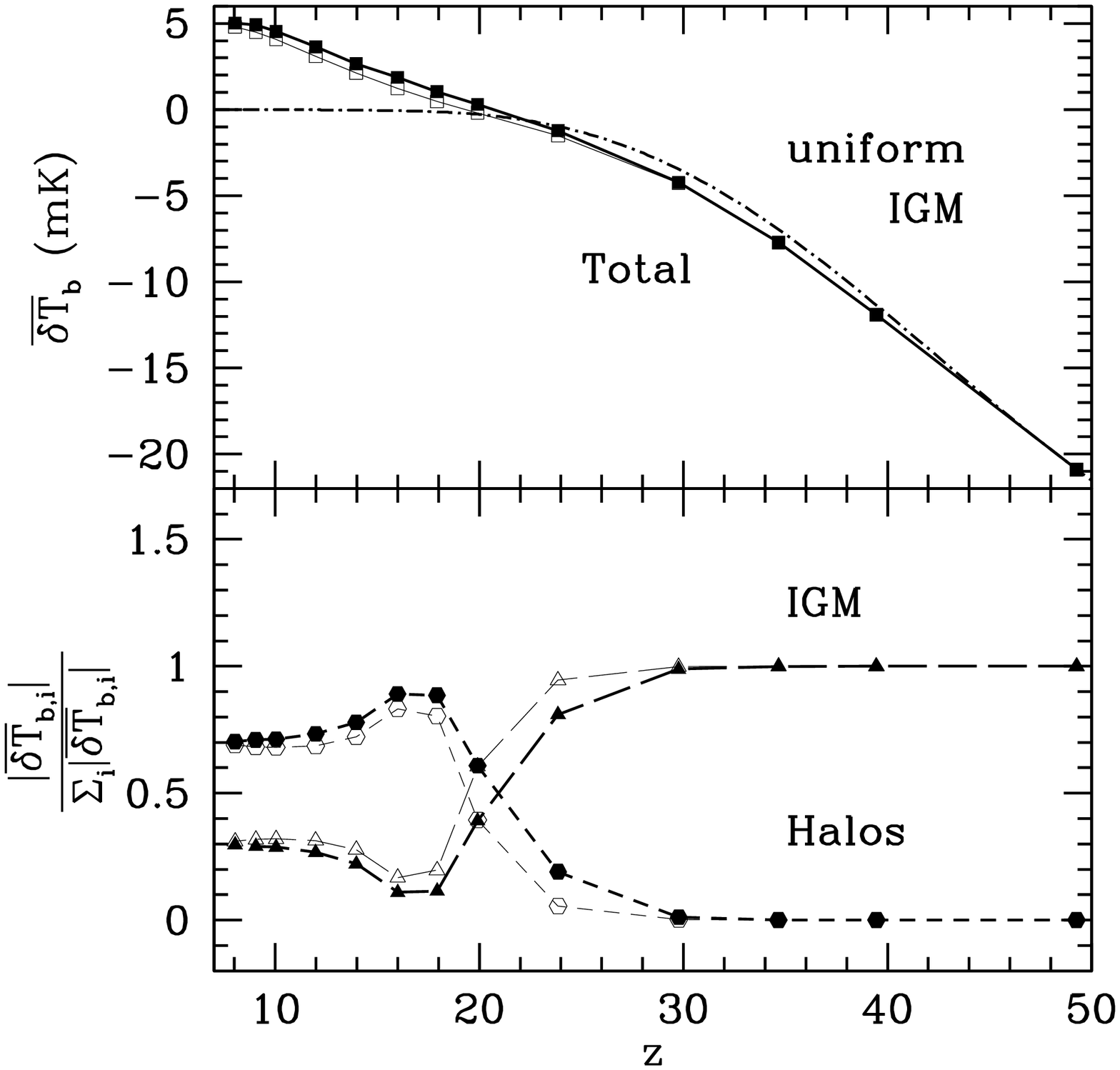}{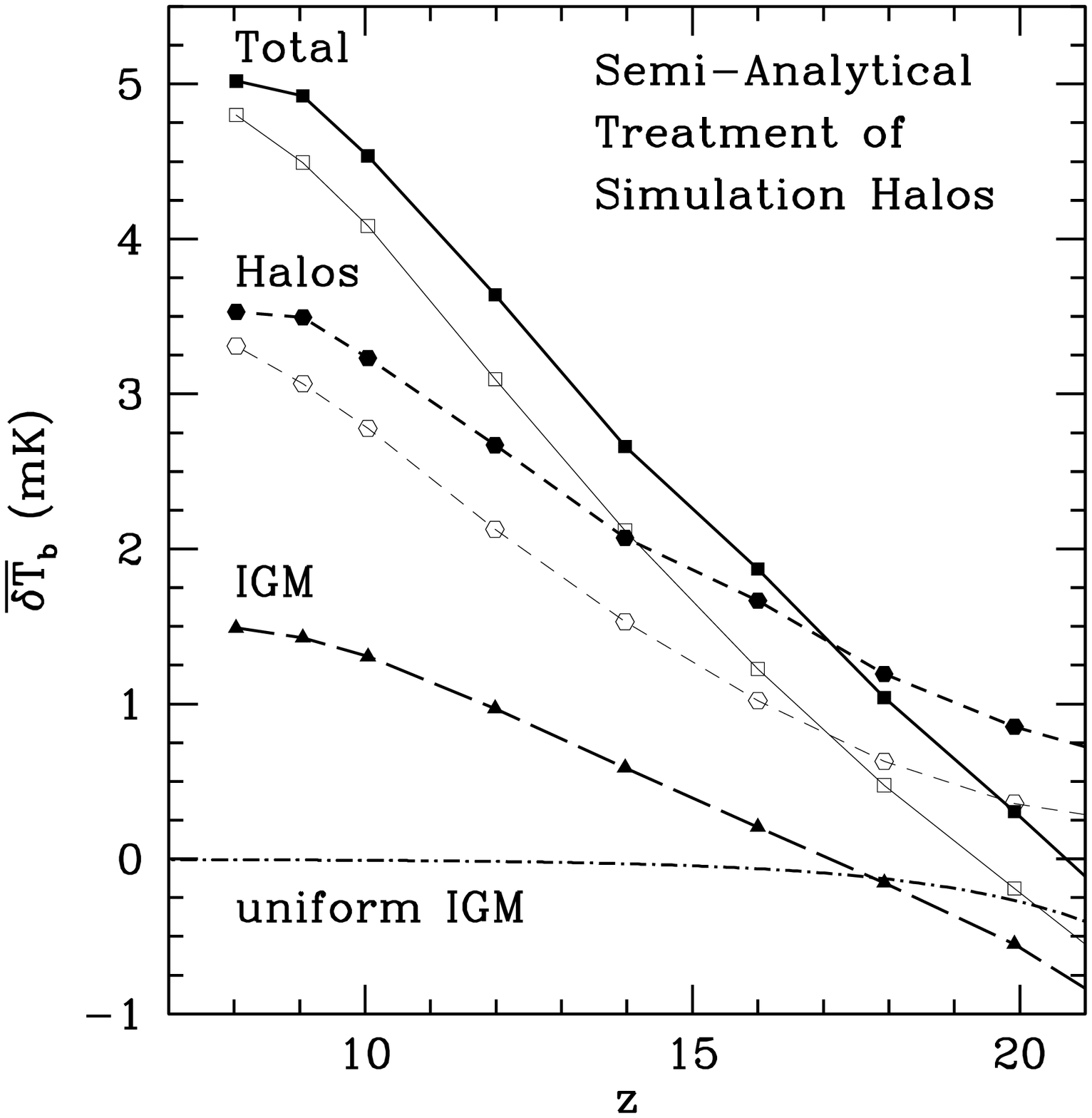}
\caption{ Semi-analytical minihalo signal vs. IGM
signal. The 21-cm flux from each halo in the simulation is found by modeling
the internal structure and 21-cm line transfer through individual
halos as described by ISFM (\S~\ref{sub:Improvements}), to calculate
the halo 21-cm signal from each halo more accurately. Same notation as in
Figure~\ref{fig-0.25-numerical}. The semi-analytical 21-cm minihalo
emission is higher than
the raw simulated minihalo signal in
Figure~\ref{fig-0.25-numerical}. The 
IGM signal remains the same. The raw minihalo and total signals
plotted in
Figure~\ref{fig-0.25-numerical} (thin lines and open symbols) are also shown for comparison.
}
\label{fig-0.25-improved}
\end{figure*}

\subsubsection{Semi-Analytical Calculation of the Halo Contribution}
\label{sub:Improvements}

Our numerical simulations have sufficiently high resolution to find all halos
in the computational box and the large-scale structure formation shocks, but
not to resolve the internal structure of the minihalos themselves. However,
as ISFM have shown, in order to obtain the correct 21-cm signal from minihalos
one needs to do a full radiative transfer calculation through each
individual minihalo density profile since, unlike the IGM
gas, minihalos have a non-negligible optical depth at the 21-cm line.
Hence, we can refine our estimate of the minihalo contribution to
the total 21-cm signal by combining our numerical halo catalogers with the
semi-analytic calculation of individual minihalo contribution as found by ISFM. 
In their approach, as described in \S~\ref{sub:minihalos},
the gas density of each minihalo is assumed to follow a TIS
profile \citep{2001MNRAS.325..468I}, radiative
transfer calculations are performed to determine the $\delta T_{b}$
for different impact parameters, 
and, finally, the face-averaged $\delta T_{b}$ is calculated 
(see ISFM, for details).
The halo mass function, $dn/dM$, is provided by the halo catalogue
we construct from the simulation. Each individual halo contribution, 
$\Delta\nu_{{\rm eff}}\delta T_{b,\,\nu_{0}}A$, depends on its mass and
redshift of formation (ISFM). Once we calculate
$\Delta\nu_{{\rm eff}}\delta T_{b,\,\nu_{0}}A$, we then obtain the
halo contribution using equation~(\ref{eq-dTb}).

\section{Results}
\label{sec:Result}

\subsection{Numerical 21-cm Brightness Temperature from Minihalos vs. IGM}
\label{sub:result-numerical}

In this section, we describe the results from our simulations. In
Figure~\ref{maps} we show (unfiltered) maps of the differential brightness
temperature obtained directly from our numerical data for our
highest-resolution simulation (C4), as
described in \S~\ref{sec:Calculation}. We show the total signal, as well as
the separate contributions from minihalos and IGM, derived as we described in
\S~\ref{sec:Numerical-Simulations}, at redshifts $z=30$, 20, and 10. 
At $z=30$, the earliest redshift shown (top row), most of the diffuse IGM gas
is still in the quasi-linear regime and cold, thus largely in absorption
against the CMB. At redshift $z=20$ (middle row), the diffuse gas is still
largely in absorption, while the (relatively few) halos that have
already collapsed are strongly in emission. The combination of the two
contributions creates a complex, patchy emission/absorption map, and 
absorption and emission partially cancel each other
in the total mean signal. Finally, at $z=10$ (bottom row),
including its diffuse component, gas heated above $T_{\rm CMB}$ is
widespread leading to a net emission against
the CMB. The bulk of this 21-cm emission comes from the high-density knots and
filaments. Although both the halo and IGM contributions come from roughly the
same regions, the minihalo emission is significantly more clustered, while the
IGM emission is quite diffuse.

In Figure~\ref{fig-accudelT}, we have plotted the volume-weighted
probability distribution functions (PDFs) for the gas density
$(1+\delta)$ and the differential brightness temperature contributions
$\delta T_b$ as functions of each other. The PDFs for gas density show
that, while the highest overdensities ($\delta \gtrsim 30$) are
typically found inside minihalos and the lower overdensities ($\delta
\lesssim 30$) and underdensities ($\delta \lesssim 0$) are typically
associated with the IGM, there is some overlap of the distributions
for these two components. A small fraction of the volume contains
lower density minihalo gas and higher density IGM gas. However, the
cumulative distributions show that these volumes hardly affect the
total mean brightness temperature contributed by each
component. Similarly, the PDFs for the brightness temperature show
that, while the volume which contributes the highest brightness
temperatures is predominantly inside minihalos and that which
contributes the lower brightness temperatures is predominantly located
in the IGM, there is, once again, some overlap of the PDFs. A small
part of the IGM volume exhibits high brightness temperature, while a
small part of the minihalo volume shows low brightness
temperature. Once again, however, the cumulative distributions show
that these regions hardly affect the total mean brightness
temperatures contributed by each component.

In Figure~\ref{fig-0.25-numerical}, we quantify the relative
contributions of the minihalos and diffuse IGM to the total mean 21-cm
signal averaged over the whole computational box and their
evolution. The evolution roughly 
follows the naive analytical estimates, as was shown in Figure~\ref{fig-anal}.
The total signal is deep in absorption, with $\delta T_b<-10$ mK at $z>37$.
The 21-cm signal is completely dominated by the IGM contribution at
this stage. The
absorption signal follows the analytical prediction for the
unperturbed universe in \S~\ref{sub:unperturbed}
well, since the density
fluctuations are still small and the uniform-density assumption is reasonably
accurate. The absorption continually decreases as significant
nonlinear structures 
start forming and portions of the gas became heated due to this structure
formation. The net signal goes into emission after redshift $z\sim 20$,
reaching up to $\sim 5$ mK by $z\approx 8$. The emission signal at
$z<18$ is due to both 
collapsed halos and the clumpy, hot IGM gas. In terms of their relative
contributions, the minihalos dominate over the diffuse IGM at
all times when the overall signal is in emission, below $z=18$.  
We find that the relative contributions to the total signal,
$\left|\overline{\delta T}_{b,\,
  j}\right|/\left(\left|\overline{\delta T}_{b,\,{\rm 
    halo}}\right|+\left|\overline{\delta T}_{b,\,{\rm IGM}}\right|\right)$ 
where $j$ means either ``halo'' or ``IGM,'' is nearly constant
over two different redshift regimes: for $z>20$, 
$\left|\overline{\delta T}_{b,\,{\rm
    IGM}}\right|/\left(\left|\overline{\delta T}_{b,\,{\rm 
    halo}}\right|+\left|\overline{\delta T}_{b,\,{\rm
    IGM}}\right|\right)\approx1$, 
while for $z<16$, $\left|\overline{\delta T}_{b,\,{\rm
    halo}}\right|/\left(\left|\overline{\delta T}_{b,\,{\rm
    halo}}\right|+\left|\overline{\delta T}_{b,\,{\rm
    IGM}}\right|\right)\approx\,0.7$. In the transition region, $16
\lesssim z \lesssim 20$ the
relative contributions exhibit more complex behavior, approximately
canceling each other
out, resulting in a total signal which is close to zero.

\subsection{Refined Estimate of the Simulated Minihalo 21-cm Signal}
\label{sub:result-improved}

As we discussed in \S~\ref{sub:Improvements}, we can improve our purely
numerical estimate of the minihalo 21-cm signal by replacing each halo's
flux with the value obtained by detailed radiative transfer calculations. 
We obtain the total minihalo signal from equation~(\ref{eq-dTb}), with
the theoretical mass function $dn/dM$ replaced by the 
actual, numerical halo catalogue obtained
from our simulations, and the individual minihalo contributions,
$\Delta\nu_{{\rm eff}}\delta T_{b,\,\nu_{0}}A$, calculated by modeling each
halo as a TIS.

\begin{figure}[t]
\plotone{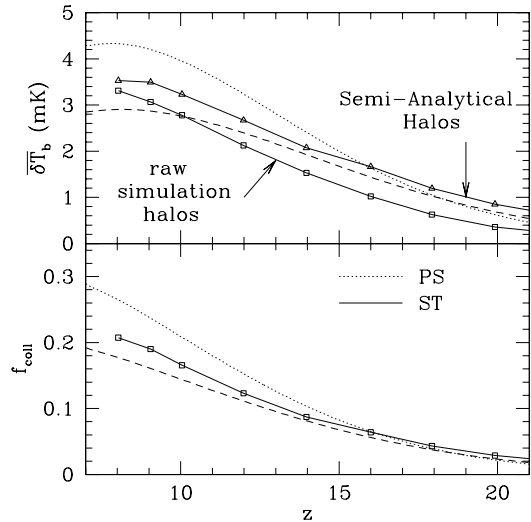}
\caption{ A comparison of analytical and
  numerical minihalo results.
  (a)(top) Differential brightness
  temperature of the 21-cm signal from minihalos for semi-analytical model
  (dotted: with Press-Schechter mass function; dashed: with Sheth-Tormen mass
  function), simulation C4 numerical result (squares) and
  semi-analytical calculation (\S~\ref{sub:Improvements}) based upon
  simulation C4 mass function (triangles). (b)(bottom) Minihalo
  collapse fraction from simulation C4 (squares) and analytical mass
  functions (line types follow those of the top panel).}
\label{fig-conv_halo}
\end{figure}

\begin{figure}[t]
\plotone{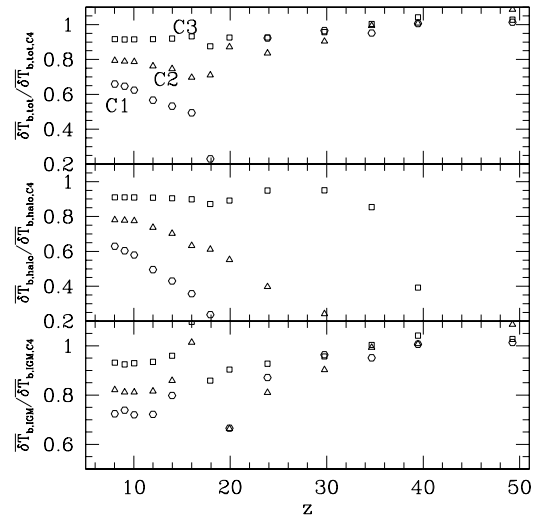}
\caption{
Numerical resolution convergence results.
 Mean differential brightness temperature signals
for simulations C1 (squares), C2 (triangles) and C3 (circles) in units of the
corresponding signal obtained from our highest resolution simulation,
C4. Shown are 21-cm signals from the total (top 
panel), halos only (middle panel) and IGM only (bottom panel).}
\label{fig-conv}
\end{figure}

We find that the resulting 21-cm signal from halos is stronger than
the ``raw'' numerical signal
obtained directly from the simulated halos and dominates the overall emission
signal even more (Fig.~\ref{fig-0.25-improved}). This is despite the fact
that our consideration of the more centrally-concentrated analytical
density profiles 
increases the optical depth of each halo. We attribute this
non-intuitive behavior to the fact that the density profiles of the minihalos
found in our simulations are not fully resolved. By modeling the halo
density profiles in detail, the local density inside halos is boosted, which
significantly increases the coupling constant $y_{c}$, which, in turn,
increases the total emission signal, even though the optical depth through
each halo also increases simultaneously. Note that we use the same
population of halos
for both estimates, and only the internal halo properties are
modified.

In Figure~\ref{fig-conv_halo}, we show the total minihalo collapsed fraction
obtained from the simulations
compared to that from the theoretical PS and ST halo mass
functions. We also show the 
minihalo contribution to the total differential brightness temperature
signal. We see that the collapsed fraction in minihalos in our
simulation roughly agrees with
the analytical predictions, mostly lying between the PS and ST results. On
the other hand, the minihalo contribution to the total 21-cm background
obtained directly from the simulation is below the theoretical predictions
based on either PS or ST mass functions. The agreement is restored, however,
when we replace each minihalo contribution to the total flux by its
analytically-modeled value. This, once again, underscores the importance of
resolving the internal halo structure for correct predictions of their 21-cm
emission. 

\subsection{Numerical Convergence}
\label{conv_sect}
We now compare cases C1, C2, C3, and C4 to check the robustness of our results
with respect to numerical convergence. In Figure~\ref{fig-conv}, we show the 
differential brightness temperature signals for 
our three lower-resolution simulations, C1, C2 and C3, in
terms of the signal obtained from our highest resolution
simulation, C4. We show the total signal, as
well as each separate contribution, from either the halos or the IGM gas.
At $z>20$ most of the gas density fluctuations are still linear, and a change
in the resolution barely affects the results. Thus a modest-resolution
simulation, or even the analytical estimate for an unperturbed IGM, 
suffices to obtain reliable
results. In contrast, at lower redshifts ($z<20$) the results depend strongly
on the resolution. The low-resolution simulations C1 and C2 underestimate the
resulting 21-cm signal significantly, by factors of up to a few. The
results from 
these simulations improve somewhat at lower redshifts, below $z=10$, but
results are still below the ones from C4 by $\sim30-50\%$ and
$\sim20\%$ for simulations C1 and C2, respectively. This is true for either
the minihalo, IGM or the total signal. The results from our medium-resolution
simulation C3, on the other hand, are much closer to the ones from the
high-resolution simulation C4, with the two generally agreeing to better than
10\%. This indicates convergence of our results to within a few per
cent for simulation 
C4. Such behavior could be na\"ively expected, since at $z<20$ non-linear
structures, both collapsed halos and mildly nonlinear, shocked IGM gas, form
in abundance at the scales we are investigating, and thus high resolution is
required to resolve these properly, as our simulations confirm. 

The relative contributions of the minihalo and the IGM signals, on
the other hand, show a more robust convergence. In all cases of different
resolutions, we find that the minihalo signal dominates the IGM signal
at $z<20,$ while the IGM signal dominates the minihalo signal at
$z>20$. For the purely numerical result, the relative contribution of minihalos
to the emission signal is about 70\% at $z<~15$, peaks to 100\% at
$z\approx18$, and drops to 50\% at $z\approx20$. The exact value
of the transition redshift varies slightly with resolution. For $z<14$,
minihalos contribute $\sim70\%$ of the emission signal. For the case of 
semi-analytical calculation of the minihalo contribution based on the
simulated halo catalogues, the relative contribution to the emission signal is
slightly higher, $\sim75\%$.

\section{Conclusions}
\label{sec:Conclusion}

\begin{figure}[t]
\plotone{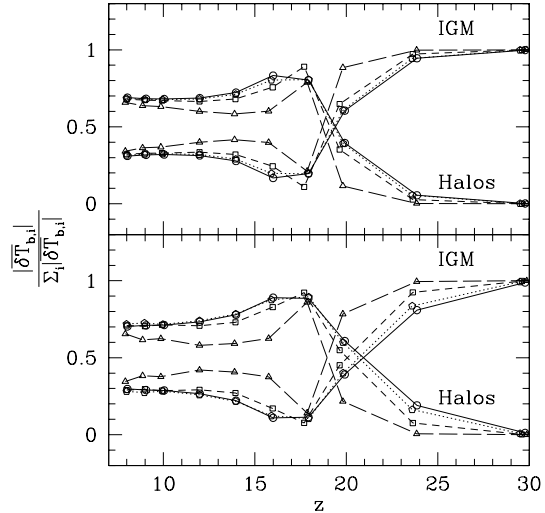}
\caption{
Numerical resolution convergence results. Relative contributions of
minihalos and diffuse 
  IGM gas to the total 21-cm background. The top panel shows the results
  obtained directly from simulations (C1: triangle, long-dashed; C2:
square, short-dashed; C3: pentagon, dotted; C4: circle, solid). 
The bottom panel shows the results which were semi-analytically
refined (\S~\ref{sub:Improvements}; point- and line-types follow those
of the top panel). }
\label{fig-conv_rel}
\end{figure}

We have run a set of cosmological N-body and hydrodynamic simulations of the
evolution of dark matter and baryonic gas at high redshift ($6<z<100$). With
the assumption that radiative feedback effects from the first light sources
are negligible, we calculated the mean differential brightness
temperature of the redshifted 21-cm background at each redshift.
The mean global signal is in absorption against the CMB
above $z\sim20$ and in overall emission below $z\sim18$. At $z>20$, the
density fluctuations of the IGM gas are largely linear, and their absorption
signal is well approximated by the one that results from assuming uniform gas
at the mean adiabatically-cooled IGM temperature. At $z<20$, nonlinear
structures become common, both minihalos and clumpy, hot, mildly nonlinear
IGM, resulting in an overall emission at 21-cm with differential brightness
temperature of order a few mK.

By identifying the halos in our simulations, we were able to separate and
compare the relative contributions of the halos and the IGM gas to the total
signal. 
We find that the emission from minihalos dominates over that from the
IGM outside minihalos, for $z\lesssim 20$. In particular,
the emission from minihalos contributes about $70\,-\,75\,\%$
of the total emission signal at $z<~17$, peaking at 100\% at
$z\approx18$, and balancing the absorption by the IGM gas at $z\approx20$.
In contrast, the absorption by cold IGM gas dominates the total signal for
$z>20$. 

These results appear to contradict the suggestion by
\citet{2004ApJ...611..642F}, that the 21-cm emission signal would be
dominated by the contribution of shocked gas in the diffuse IGM.
They used the Press-Schechter formalism to estimate the fraction of
the IGM outside of minihalos, which is shock-heated, by adopting a
spherical infall model for the growth of density fluctuations and
assuming that all gas inside the turn-around radius is
shock-heated. This method is apparently not accurate enough to
describe the filamentary nature of structure formation in the IGM. 

On the other hand, our results are consistent with the analytical
estimates of the mean 21-cm emission signal from minihalos by
ISFM. This indicates that the statistical prediction of the collapsed
and virialized regions identified as minihalos by the Press-Schechter
formalism (or its refinement in terms of the ST formula), with virial
temperatures $T < 10^4 {\rm K}$, with halos characterized individually
by the TIS model, is a reasonably good approximation for the mean
21-cm signal for minihalos at all redshifts and a good estimator even
for the total mean signal including both minihalos and the diffuse IGM
at $z \lesssim 20$. This encourages us to believe that the angular and
spectral fluctuations in the 21-cm background predicted by ISFM based
on that model will also be borne out by future simulations involving a
much larger volume than was simulated here. The current simulation
volume is too small to be used to calculate the fluctuations in the
21-cm background because current plans for radio surveys to measure
this background involve beams which will sample much larger angular
scales ($>$ arcminutes) than are subtended by our current box ($\Delta
\theta_{\rm box} \sim 0.2' (1+z)_{20}^{0.2}$, where
$(1+z)_{z'}=(1+z)/(1+z')$) and bandwidths ($\sim $ 
MHz) which are 
too large to resolve the depth of our simulation box in redshift-space
(i.e. $\Delta \nu_{\rm box} \sim 40 {\rm kHz} (1+z)_{10}^{-1/2}
\left[L/(0.5 h^{-1} {\rm Mpc}) \right] $ ). According to ISFM and
\citet{2003MNRAS.341...81I}, the fluctuations in the 21-cm background
from minihalos are significantly enhanced by the fact that minihalos
are biased relative to the total matter density fluctuations. A larger
simulation volume than ours will also be necessary to sample this
minihalo bias in a statistically meaningful way. This bias is likely
to affect the minihalo contribution to the 21-cm background
fluctuations substantially more than it does the diffuse IGM
contribution, thereby boosting the relative importance of minihalos
over the IGM even above the ratio of their contributions to the mean signal.

We have considered the limit in which only collisional pumping is
available to decouple the spin temperature from that of the CMB, and
sources of radiative pumping have not yet emerged to compete with this
process. The possibility exists, however, that an X-ray background
built up as sources formed inside some halos, which heated the IGM while
only partially ionizing it
(e.g. \citealt{2003MNRAS.346..456O}). This heating might have boosted the
kinetic temperature of the IGM and enhanced the effect of collisional
pumping there (e.g. \citealt{2004ApJ...602....1C})\footnote{Recently, 
  \citet*{2006ApJ...637L...1K} considered the X-rays emitted by an
  early miniquasar, finding that such an X-ray source can heat the IGM
to as much as a few thousand degrees Kelvin without ionizing it. This
boosts the 21-cm signal from collisionally-decoupled gas in the
diffuse IGM significantly. Their calculations neglect the ionizing UV
radiation which might also be released by the miniquasar and its
stellar progenitor, as well as the Ly$\alpha$ pumping they might
contribute.}. Such X-ray heating would also have raised the minimum
mass of minihalos which formed thereafter, filled with their fair
share of neutral H atoms.
When stellar sources began to form and build
up the UV background at energies below the Lyman limit of hydrogen,
Ly$\alpha$ pumping could then have radiatively coupled $T_{\rm S}$ to
$T_{\rm K}$, as well. 
The same sources presumably emitted UV radiation above the H Lyman
limit, too, which ionized both the IGM and the minihalos within the
HII regions surrounding these sources
(e.g. \citealt*{2004MNRAS.348..753S}; \citealt*{ISR05};
\citealt*{ISS05}). Such HII regions would have created holes in the
21-cm background, which then originated only in the remaining neutral
regions. Minihalos could have lost the neutral hydrogen gas
responsible for their 21-cm emission, not only by ``outside-in''
photoionization by an external source, but also by ``inside-out''
photoionization by internal Pop III star formation
(e.g. \citealt{2004ApJ...613..631K}; \citealt*{2006ApJ...639..621A}). 
The $\rm H_2$
formation required for minihalos to form stars, however, is likely to have
been suppressed easily by photodissociation in the Lyman-Warner bands
by the background of UV radiation created by the very first minihalos
which formed stars, when the ionizing radiation background was still
much too low to cause reionization (\citealt*{2000ApJ...534...11H}). In
that case, most minihalos would have remained intact until they were
ionized from without.
In the future, we plan to improve upon 
the current calculation by incorporating this more complicated physics.
We also intend to run simulations with larger simulation boxes. This would
allow us to  predict the 21-cm fluctuation signal (e.g. ISFM)
and determine whether the relative contribution of minihalos to the total
signal, which we find to be about 70 -- 75 \% at $z \lesssim 20$ for the
mean signal, varies
as the mean signal fluctuates.

\acknowledgments
This work was partially supported by NASA Astrophysical Theory
Program grants NAG5-10825,
NAG5-10826, NNG04GI77G, and Texas Advanced Research Program grant
3658-0624-1999. MAA was supported by a DOE Computational
Science Graduate Fellowship.
HM is supported by NSERC. The work by DR was supported by the KOSEF grant
R01-2004-000-10005-0.

\end{document}